\documentstyle[11pt,psfig,epsfig]{article}
\begin{document}
\title{The extent of strangeness equilibration  in quark-gluon plasma}

\author{{\normalsize Dipali Pal \thanks{Acknowledges support from WIS, Israel and SINP, India.}}\\
\small {\em Department of Particle Physics, Weizmann Institute of Science, Rehovot 76100, Israel.}\\ \\
{\normalsize {Abhijit Sen, Munshi Golam Mustafa}} \\
\small {\em Theory Group, Saha Institute of Nuclear Physics, 1/AF, Bidhan Nagar, Calcutta - 700064, India}\\ \\
{\normalsize Dinesh Kumar Srivastava} \\
\small {\em Variable Energy Cyclotron Centre, 1/AF, Bidhan Nagar, Calcutta - 700064, India}}
\maketitle

\begin{abstract} 
The evolution and production of strangeness from chemically 
equilibrating and transversely expanding quark gluon plasma which 
may be formed in the wake of relativistic heavy ion collisions is 
studied with initial conditions obtained from the Self Screened Parton 
Cascade (SSPC) model. The extent of partonic equilibration increases almost
linearly with the square of the initial energy density, which can then 
be scaled with number of participants. 
\end{abstract} 


\section{Introduction}
Strangeness enhancement is one of the robust signatures of quark - 
hadron phase transition during the ultra relativistic heavy ion 
collisions~\cite{sig,sig3}. In heavy ion collisions strangeness 
is produced abundantly through the partonic interactions if the 
temperature $T \ge 200$ MeV, the mass threshold of this semi-heavy 
flavour. The extent of its equilibration would however depend upon 
the initial conditions and the life time of the hot deconfined 
phase. It has recently been shown that the chemical equilibration 
of the light flavours and the gluons slows down due to the radial 
expansion~\cite{chem2,duncan}. It should then be expected that 
the extent of strangeness equilibration can also be affected if 
the radial expansion of the plasma is included. An early work in 
this direction used the initial conditions obtained from the HIJING 
model and considered only a longitudinal expansion~\cite{levai}. 
In the present work, we closely follow this treatment with the 
initial conditions obtained from Self Screened Parton Cascade 
(SSPC) model ~\cite{sspc} and extend it to include transverse 
expansion~\cite{pal} as well.
In the next section we briefly describe the hydrodynamic and chemical 
evolution of the plasma in a (1+1) dimensional longitudinal expansion 
and a (3+1) dimensional transverse expansion. A brief summary is given 
in Sec. 3.

\section{Hydrodynamic Expansion and Chemical Equilibration}

\subsection{Basic equations}

We start with the assumption that the system achieves a kinetic 
equilibrium by the time $\tau_i$ and the chemical equilibration is 
assumed to proceed via gluon multiplication process 
($ gg \rightarrow ggg$) and quark production process 
($ gg \rightarrow q {\bar q}$). The expansion of the system is now 
controlled by the equation for conservation of energy and momentum 
of an ideal fluid:
\begin{equation}
\partial_\mu T^{\mu \nu}=0 \; , \qquad
 T^{\mu \nu}=(\varepsilon+P) u^\mu u^\nu + P g^{\mu \nu} \, ,
\label{hydro}
\end{equation}
where $\varepsilon$ is the energy density and $P$ is the pressure 
measured in the rest frame of the fluid~\cite{pal}. The four-velocity 
vector $u^\mu$ of the fluid satisfies the constraint $u^2=-1$.                                                                       
We assume that the distribution functions for partons can be scaled
through equilibrium distributions as
\begin{equation}
f_j(E_j,\lambda_j)= \lambda_{j} {\tilde f}_j(E_j) \ \ , \label{befd}
\end{equation}
where ${\tilde f}_j(E_j)=({e^{\beta{E_j}}\mp 1})^{-1}$
is the BE (FD) distribution for gluons
(quarks), and $\lambda_j$ ($j=g, \ u, \ d, \ s$)  are the 
nonequilibrium fugacities, $E_j=\sqrt{p_j^2+m_j^2}$, and $m_j$ is the 
mass of the parton.                                    
We solve the hydrodynamic equations (\ref{hydro}) with the assumption
that the system undergoes a boost invariant longitudinal expansion 
along the $z$-axis and a cylindrically symmetric transverse expansion 
\cite{vesa}. It is then sufficient to solve the problem for $z=0$. 

The master equations governing the chemical equilibration for the 
dominant chemical reactions are 

\begin{eqnarray}
\partial_\mu (n_g u^\mu)&=&(R_{2 \rightarrow 3} -R_{3 \rightarrow 2})
                    - \sum_i \left ( R_{g \rightarrow i}
                       - R_{i \rightarrow g} \right ) \, , \nonumber\\
\partial_\mu (n_i u^\mu)&=&\partial_\mu (n_{\bar{i}} u^\mu)
                     =  R_{g \rightarrow i}
                       - R_{i \rightarrow g},
\label{master1}
\end{eqnarray}                                    
in an obvious notation.
The gain and loss term for the gluon fusion
($gg\leftrightarrow i{\bar i}$) and gluon multiplication 
($gg\leftrightarrow ggg$) processes can be written as 

\begin{eqnarray}
R_{g \rightarrow i}-R_{i \rightarrow g} &=&R_2^i n_g                    
\left(1-\frac{\lambda_i^2} {\lambda_g^2}\right). \label{fg}
\end{eqnarray}  
\begin{eqnarray}  
R_{2 \rightarrow 3} -R_{3 \rightarrow 2}&=&R_3 n_g \left(1-\lambda_g\right)
. \label{gm1}
\end{eqnarray}  
Using these rate equations, Eq.(\ref{master1}) can now be simplified
for (3+1) and (1+1) dimension accordingly~\cite{pal}. For solving the 
eqs.(\ref{hydro}) and (\ref{master1}) numerically, we use the initial 
conditions obtained from SSPC model~\cite{sspc}. In calculating the
thermally averaged and velocity weighted rates, $R_2^i$ and $R_3$, 
we have also considered the temperature dependent coupling 
constant, however, for details of the calculation see Ref.~\cite{pal}.

\newpage

\subsection{Results and Discussions}
\vspace{-0.2in}
\begin{figure}[htbp]
\centerline{{\epsfxsize=5cm {\epsfbox{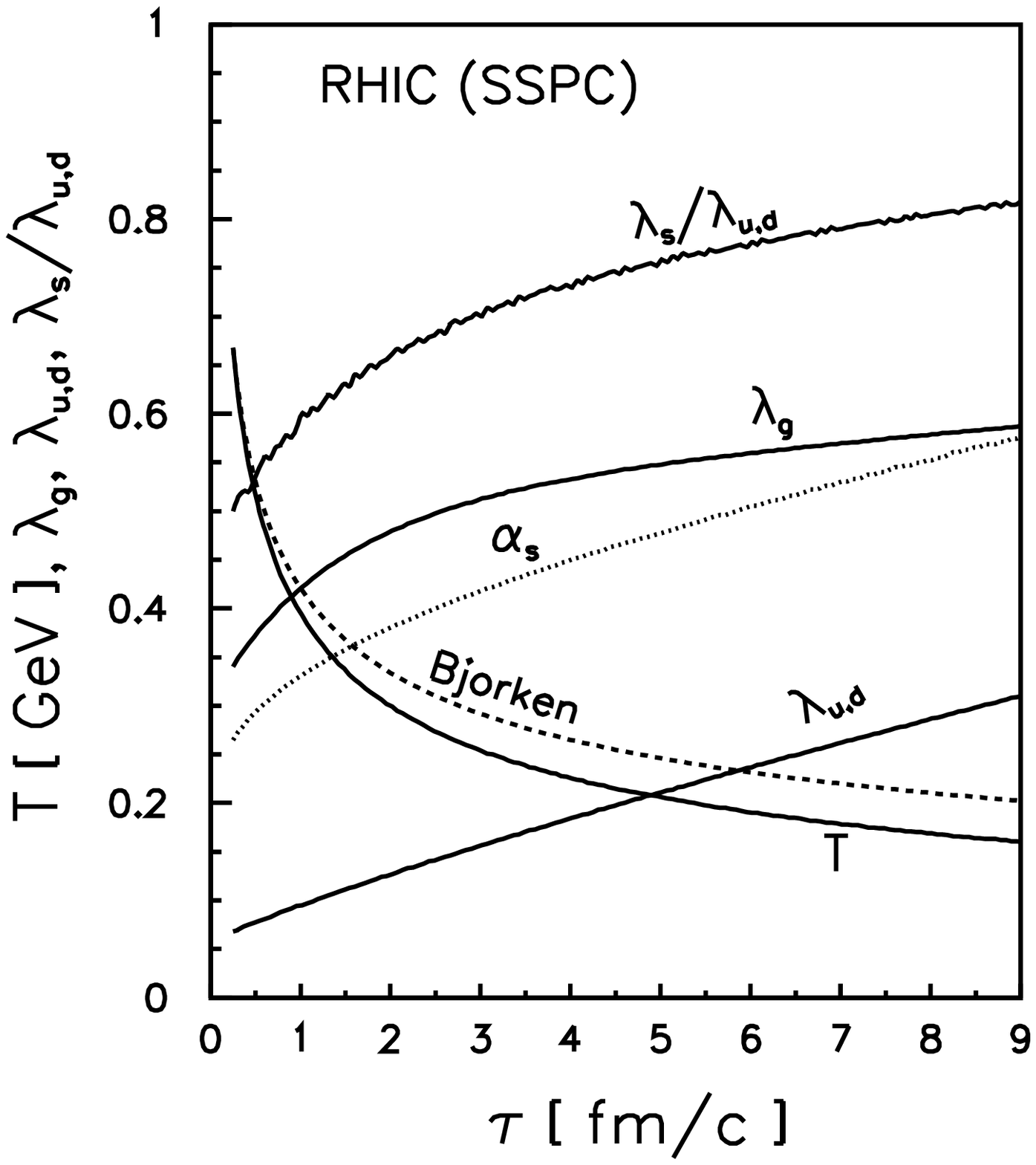}}}
{\epsfxsize=5cm {\epsfbox{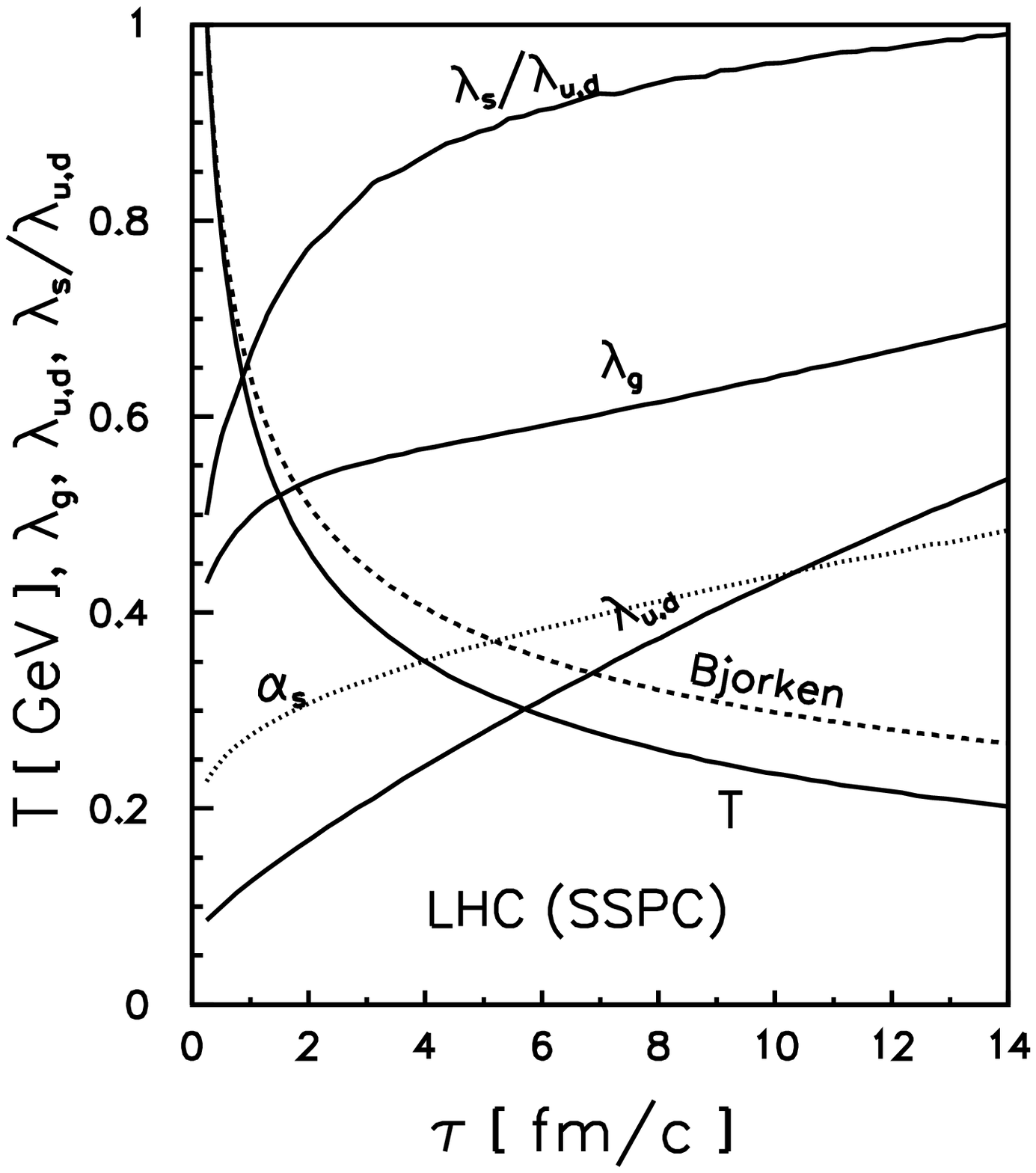}}}}
\vspace{-0.3in}
\caption{Variation of temperature, coupling constant, gluon and quark 
fugacities with proper time for (1+1) dimensional hydrodynamic 
expansion with SSPC initial conditions for RHIC (left) and LHC (right)
energies.}
\label{fig:long}
\end{figure}                                                           

In figure~\ref{fig:long} we show our results for the longitudinal expansion for 
RHIC and LHC energies. We do note that the plasma is not fully 
chemically equilibrated at either RHIC or LHC energies. As the 
additional parton production consumes energy, the temperature of the 
partonic plasma is found to be reduced considerably faster than the 
ideal Bjorken's scaling ($T=T_0 (\tau_0/\tau)^{1/3}$, $T_0$ and 
$\tau_0$, respectively, are initial temperature and time of the parton 
gas) represented by the dashed line. We further note that the extent 
of equilibration for the strange quarks in comparison to that for the 
light quarks ($\lambda_s/\lambda_{u,d}$) rises rapidly and once the 
temperature falls below about $300$ MeV ($\sim \ 2m_s$ ) it gets more 
or less frozen by this time.      

\begin{figure}[htbp]
\vspace{-0.2in}
\centerline{{\epsfxsize=5cm {\epsfbox{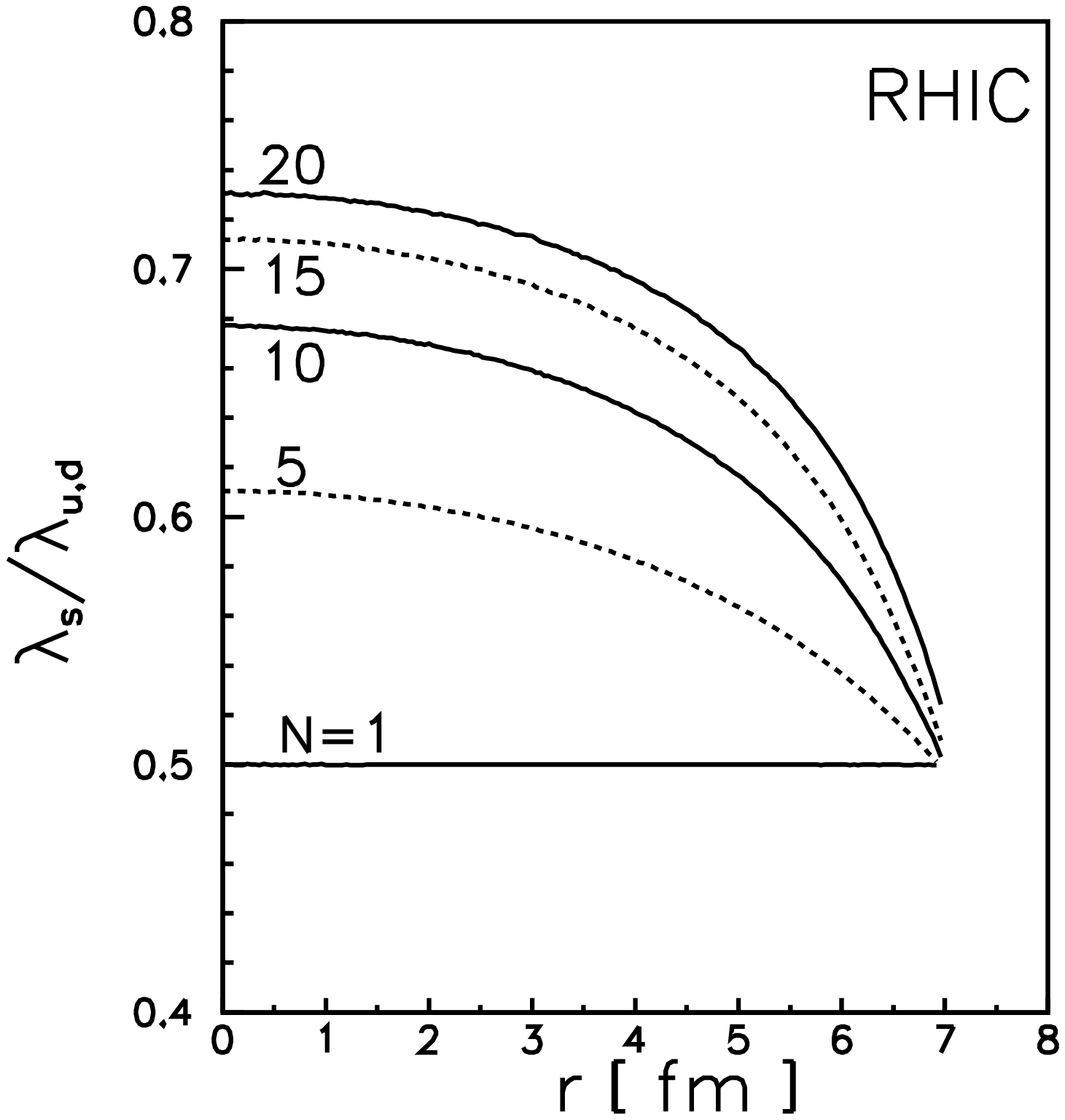}}}
{\epsfxsize=5cm {\epsfbox{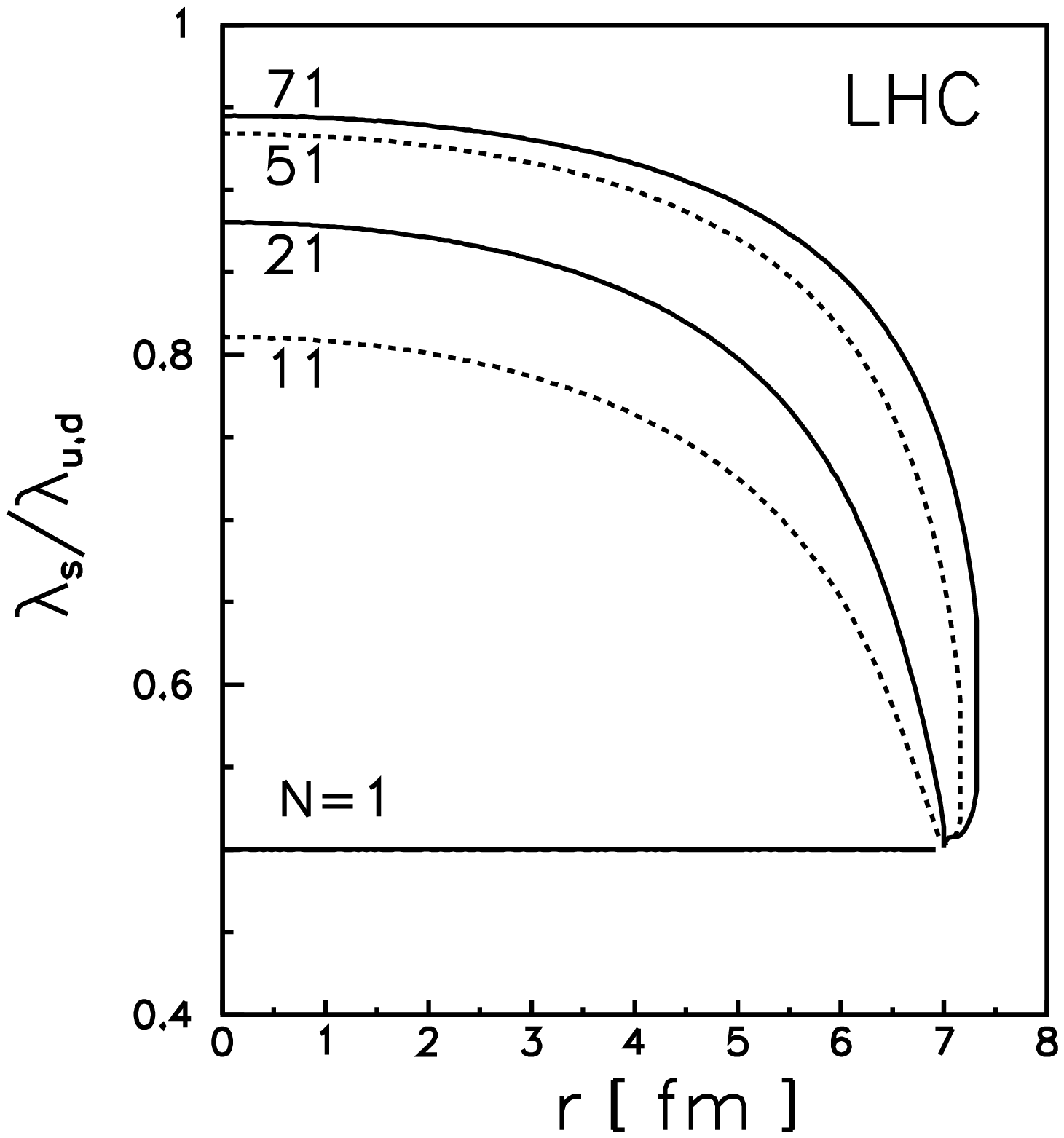}}}}
\vspace{-0.3in}
\caption{Variation of strange to non-strange quark fugacity ratio with 
the transverse radius for RHIC (left) and LHC (right) 
energies at different times.}
\label{fig:rlrat}
\end{figure} 

In figure~\ref{fig:rlrat}, we present our results for the radial 
variation of $\lambda_s/\lambda_{u,d}$ for RHIC and LHC energies, 
respectively, for various times along the constant energy density 
contours with $\tau =N\tau_0$. Here $N$ is defined~\cite{chem2,chem3} 
through $\varepsilon(r,\tau)=\varepsilon(r=0,\tau_0)/N^{4/3}$.          
We see that the extent of strangeness equilibration attains its
highest value near $r=0$ and rises rapidly first and only slowly 
later in time, though the individual variation of differnet 
$\lambda$'s are somewhat interesting~\cite{pal}.  
Here also the plasma is not fully equilibrated as before.

\begin{figure}[htbp]
\vspace{-0.2in}
\centerline{{\epsfxsize=5cm {\epsfbox{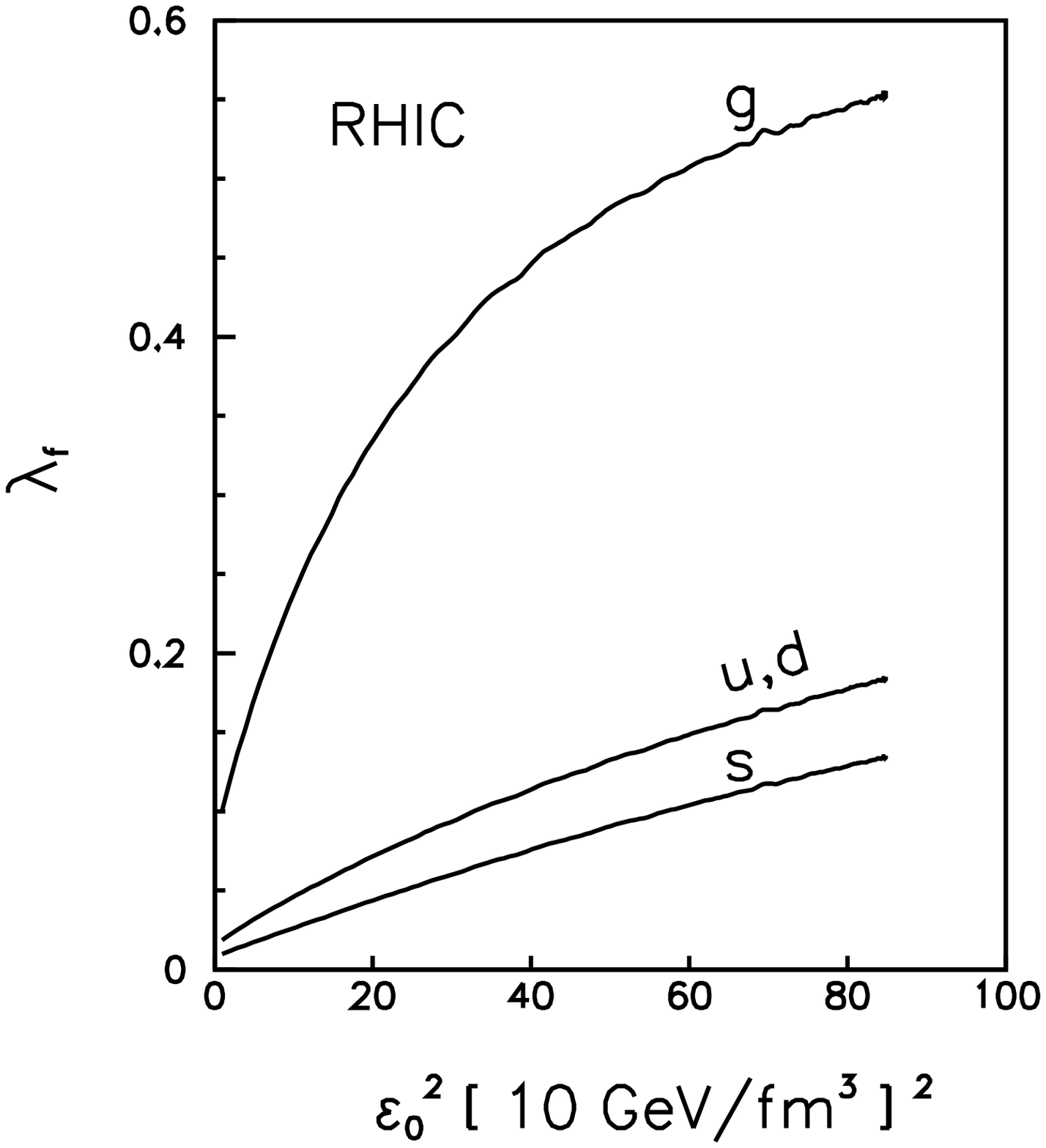}}}
{\epsfxsize=5cm {\epsfbox{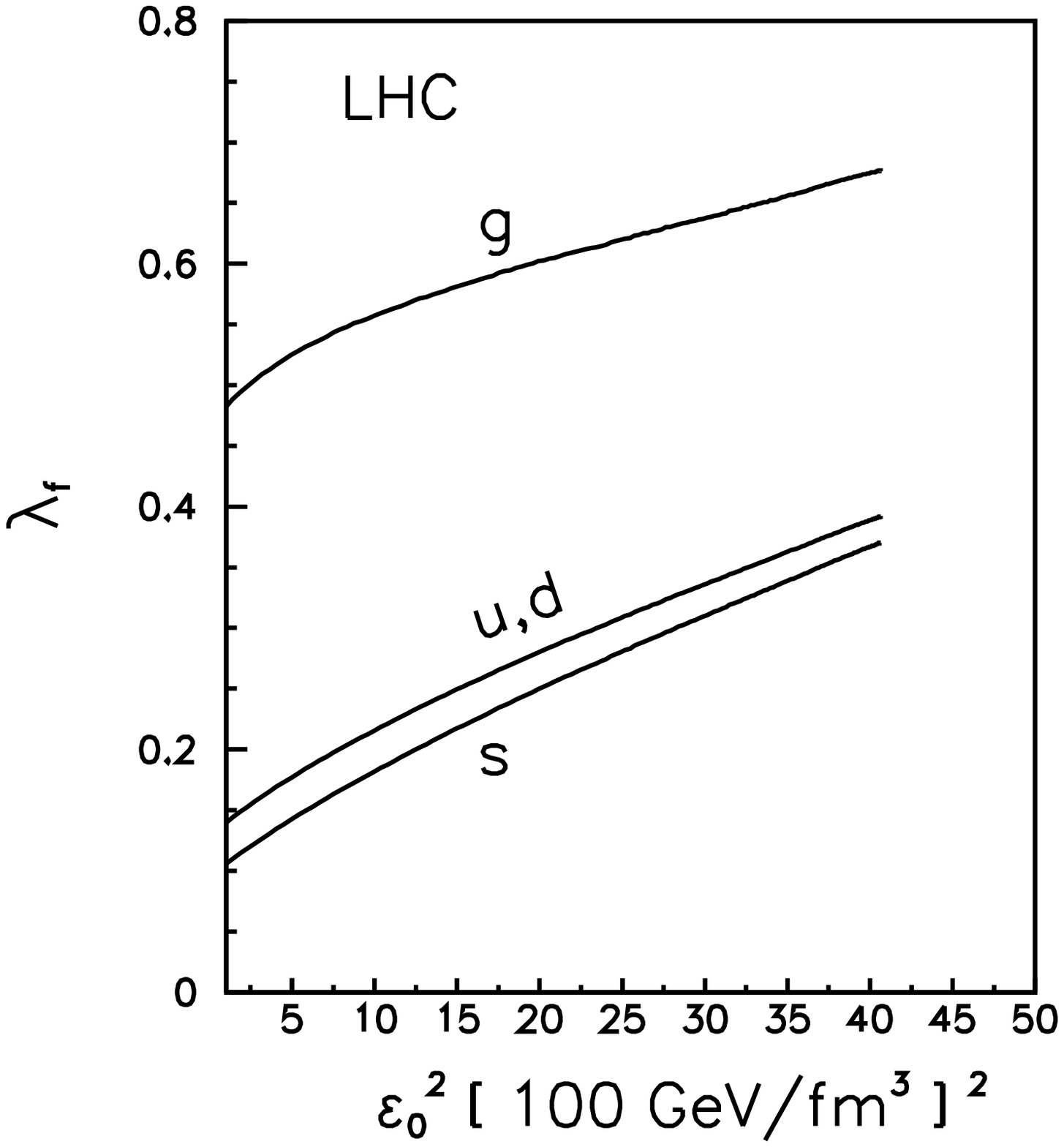}}}}
\vspace{-0.3in}
\caption{Variation of the final fugacities with the initial energy 
density for RHIC (left) and LHC (right) energies.}
\label{fig:rlene}
\end{figure}

The radial variation of the final fugacities with the initial energy 
density for RHIC and LHC energies are displayed in 
figure~\ref{fig:rlene}. It 
is interesting to note that once the energy density is beyond about 
20--40 GeV, the final fugacities for all the partons increase almost 
linearly with the square of the energy density~\cite{pal} as obtained at 
the CERN SPS~\cite{jean} energies. 

\section{Summary}

We have studied the evolution and production of strangeness from an 
equilibrating and transversely expanding quark gluon plasma. Initial 
conditions are obtained from SSPC model. We find that most of the 
strange quarks are produced when the temperature is still more than 
about 300 MeV ($\sim 2m_s$) and are not fully equilibrated. 
We also find  approximately that the number of strange quarks produced 
(extent of strangeness equilibration) 
rises linearly with the square of 
the initial energy density within our approach. This may help us to obtain
the scaling of the initial energy density with the number of participants.

\end{document}